\documentclass[%
 reprint,
superscriptaddress,
 amsmath,amssymb,
 aps,
 pra,
]{revtex4-1}

\usepackage{graphicx}
\usepackage{color}
\usepackage{dcolumn}
\usepackage{bm}

\begin{document}


\title{Using surface lattice resonances to engineer nonlinear optical processes in metal nanoparticle arrays }

\author{Mikko J. Huttunen}
\email{mikkojhuttunen@gmail.com}
\affiliation{Department of Physics, University of Ottawa, Ottawa, Ontario K1N 6N5, Canada\\}%
\affiliation{Laboratory of Photonics, Tampere University of Technology, FI-33100 Tampere, Finland}

\author{Payman Rasekh}%
\affiliation{Department of Physics,  University of Ottawa, Ottawa, Ontario K1N 6N5, Canada\\}%
\affiliation{School of Electrical Engineering and Computer Science, University of Ottawa, Ottawa,  Ontario K1N 6N5, Canada}

\author{Robert W. Boyd}
\affiliation{Department of Physics,  University of Ottawa, Ottawa, Ontario K1N 6N5, Canada\\}%
\affiliation{School of Electrical Engineering and Computer Science, University of Ottawa, Ottawa,  Ontario K1N 6N5, Canada}
\affiliation{The Institute of Optics and Department of Physics and Astronomy, University of Rochester, Rochester, New York 14627, USA}

\author{Ksenia Dolgaleva}
\affiliation{Department of Physics,  University of Ottawa, Ottawa, Ontario K1N 6N5, Canada\\}%
\affiliation{School of Electrical Engineering and Computer Science, University of Ottawa, Ottawa,  Ontario K1N 6N5, Canada}

\date{\today}

\begin{abstract}
Collective responses of localized surface plasmon resonances, known as surface lattice resonances (SLRs) in metal nanoparticle arrays, can lead to high quality factors ($\sim$100), large local-field enhancements and strong light-matter interactions. SLRs have found many applications in linear optics, but little work of the influence of SLRs on nonlinear optics has been reported. Here we show how SLRs could be utilized to enhance nonlinear optical interactions. We devote special attention to the sum-frequency, difference-frequency, and third-harmonic generation processes because of their potential for the realization of novel sources of light. We also demonstrate how such arrays could be engineered to enhance higher-order nonlinear optical interactions through cascaded nonlinear processes. In particular, we demonstrate how the efficiency of third-harmonic generation could be engineered via cascaded second-order responses.

\end{abstract}

\maketitle

\section{Introduction}
An important application of nonlinear optics is the development of means to create coherent light sources at frequencies other than the fundamental laser frequency. Among the specific light sources realized in such a manner are optical parametric oscillators~\cite{boyd2003nonlinear}, sources of terahertz (THz) and extreme ultraviolet radiation~\cite{corkum1993plasma}, single and entangled photon sources~\cite{kwiat1995new,howell2004realization}. Better understanding of nonlinear optical processes, as well as the search of new mechanisms of their enhancement, could lead to the development of new improved light sources advancing many sub-fields of natural science.  

Most nonlinear processes are inherently weak. Therefore, strong excitation fields are often needed to achieve nonlinear responses using traditional nonlinear crystals. One practical solution for enhancing higher-order nonlinearities are so-called cascaded processes, where considerably stronger lower-order nonlinearities are utilized to mimic higher-order nonlinear interactions~\cite{boyd2003nonlinear,chou19991, misoguti2001generation}. 

Alternatively, more efficient nonlinear processes can be realized with nonlinear optical fibers, waveguides or resonators with high Q-factor values, such as photonic crystal cavities or ring resonators~\cite{helt2012does}. Although the intrinsic material nonlinearities of fibers and waveguides are often quite weak, the long interaction lengths can lead to strong nonlinear responses due to a coherent build-up of the nonlinear optical signal during the propagation. High Q-factor value resonators work seemingly in a similar way, since the light coupled into the resonators can make several round-trips, thereby increasing the interaction length~\cite{helt2012does}. But in order to better explain the nonlinear enhancement mechanisms in resonators, a more detailed description based on local density of states is needed~\cite{helt2012does, rivoire2011multiply}. 

Recently, plasmonics has been recognized as a possible route for more efficient nonlinear optical processes~\cite{kauranen2012nonlinear}. This is mostly due to the fact that typical metals have $\sim$10$^6$ stronger nonlinearities compared to those of typical dielectric materials~\cite{bennink1999accessing}. In addition, conduction electron oscillations in metal nanoparticles can exhibit resonant behavior~\cite{barnes2003surface}. These resonances are called  localized surface plasmon resonances (LSPRs); they can greatly enhance the local fields and the occurring light-matter interactions near the particle surface. On the other hand, the LSPRs have quite low intrinsic Q-factor values ($\sim$5). A potential solution to this problem is to utilize collective responses of metal nanoparticle arrays, also known as surface lattice resonances (SLRs)~\cite{auguie2008collective}. SLRs occur when the optical path length between the neighboring particles is an integer multiple of the incident wavelength and can exhibit considerably higher Q-factor values ($\sim$100) compared to those of LSPRs. Since larger Q-factor values imply higher local fields, especially nonlinear light-matter interactions can be expected to be considerably enhanced near SLRs~\cite{Huttunen2016strong}. But so far, very little work has been conducted to systematically study the possibilities of utilizing periodic arrays of nanoparticles and SLRs for enhancing nonlinear optical processes~\cite{czaplicki2013enhancement, metzger2015strong,Czaplicki2016surface,Suchowski2017nonlinear}. 

In this Paper, we systematically study how nanoparticle arrays and SLRs could be utilized for enhancing nonlinear optical processes. For this task, we implement a nonlinear discrete-dipole approximation (DDA) approach to simulate the linear and nonlinear optical responses of the nanoparticle arrays~\cite{draine1994discrete, balla2010second}. We study several array configurations and show how processes such as second-harmonic generation (SHG), sum-frequency generation (SFG), difference-frequency generation (DFG) and third-harmonic generation (THG) can be considerably enhanced in the vicinity of SLRs. In addition, we show how two consecutive second-order processes of SHG and SFG could be utilized to give rise to a strong cascaded contribution to THG. Our results show the promise of using metal nanoparticle arrays as a highly versatile nonlinear material platform.

\section{Theory}
\subsection{Discrete-dipole approximation for the fundamental-frequency field}
We use the DDA to study the nonlinear behavior of nanoparticle arrays, since it has been previously found to be a powerful tool to understand the linear collective responses of the arrays~\cite{markel1993coupled,zou2004silver,Kataja2015,Huttunen2016strong}. The nanoparticle array is considered to be illuminated by an incident electromagnetic field oscillating at the fundamental frequency $\omega$. This field induces a dipole moment in each of the nanoparticles. Since each of these dipole moments is also affected by the scattered field due to the presence of other dipoles, it is convenient to find the self-consistent dipole moments of the nanoparticles numerically. Once the self-consistent dipole moment distribution has been found, it can be used to calculate the associated local field acting on the nanoparticles. 
Since it is the local fields that drive the atomic transitions, thereby influencing the nonlinear optical interactions~\cite{ksenia2012local}, we repeat the calculation for each frequency component $\omega_i$ of relevance for the nonlinear phenomena under study. As explained in detail in the next section, we then use these local field components to calculate the nonlinear optical response of the nanoparticle arrays following the approach described in Ref.~\cite{balla2010second}.   

We start by assuming that the incident field $\mathbf{E}_{\mathrm{inc},j}(\omega_1)$, oscillating at the frequency $\omega_1$ at the location $\mathbf{r}_{j}$ of the $j^{\mathrm{th}}$ scattering nanoparticle (or dipole), is a monochromatic plane wave of the form
\begin{equation}  \label{Eq:Einc}
\mathbf{E}_{\mathrm{inc},j}(\omega_1) = \mathbf{E}_{0} \, \mathrm{exp} ( \mathrm{i} \mathbf{k}_1 \cdot \mathbf{r}_j - \mathrm{i} \omega_1 t) \, ,
\end{equation}
where $\mathbf{E}_{0}$ is the field amplitude, $\mathbf{k}_1$ is the wave vector and $t$ is time. The incident field interacts with all other dipoles in the array and gives rise to the total field at the site of dipole $j$ in the form
\begin{equation}  \label{Eq:Etot}
\mathbf{E}_j(\omega_1) = \mathbf{E}_{\mathrm{inc},j}(\omega_1) - \sum_{k \neq j} \mathbf{A}_{jk}(\omega_1) \mathbf{p}_{k}(\omega_1) \, ,
\end{equation}
where $\mathbf{A}_{jk}(\omega_1)$ is a $3 \times 3$ matrix describing the interaction between the $j^{\mathrm{th}}$ and $k^{\mathrm{th}}$ dipoles, and $\mathbf{p}_k(\omega_1)$ is the dipole moment of the $k^{\mathrm{th}}$ dipole at frequency $\omega_1$. When the dipoles are embedded in a homogeneous medium, their interaction is governed by a tensorial free-space Green's function and can be written 
as~\cite{draine1994discrete}
\begin{align}  \label{Eq:Aij} 
\mathbf{A}_{jk} & (\omega_1)  =  \frac{\mathrm{exp}(\mathrm{i} k_1 r_{jk})}{\epsilon_0  r_{jk}}  \nonumber \\ & \times \left[ k_1^2 
(\hat{\mathbf{r}}_{jk} \hat{\mathbf{r}}_{jk} -\mathbf{I}) - \frac{1-\mathrm{i} k_1 r_{jk}}{r_{jk}^2}
(3 \hat{\mathbf{r}}_{jk} \hat{\mathbf{r}}_{jk} - \mathbf{I}) \right] . 
\end{align}
Here $\epsilon_0$ is the vacuum permittivity, $k_1=\lvert \mathbf{k}_1 \rvert = 2\pi n / \lambda_1$ is the wavenumber, $n$ is the refractive index of the surrounding medium, $\lambda_1$ is the wavelength, $r_{jk}$ is the distance between the dipoles and $\hat{\mathbf{r}}_{jk}$ is the unit vector pointing in the direction from $\mathbf{r}_j$ to $\mathbf{r}_k$. The terms $\mathbf{I}$ are $3 \times 3$ identity matrices. It is possible to define $3 \times 3$ diagonal blocks of the interaction matrix as $\mathbf{A}_{jj}(\omega_1) =\alpha_j^{-1}(\omega_1)$, where $\alpha_j(\omega_1)$ is the polarizability of the $j^{\mathrm{th}}$ dipole at the frequency $\omega_1$. We can then rewrite Eq.~(\ref{Eq:Etot}) as
\begin{equation} \label{Eq:Etot2}
\mathbf{E}_{\mathrm{inc},j}(\omega_1) = \mathbf{A}_{jj}(\omega_1) \mathbf{p}_{j}(\omega_1) + \sum_{k \neq j} \mathbf{A}_{jk}(\omega_1) \mathbf{p}_{k}(\omega_1) \, ,
\end{equation}
which can then be written as a system of $3N$ linear equations with $3N$ unknown dipole moment components as 
\begin{equation} \label{Eq:DDA_1}
\mathbf{E}_{\mathrm{inc},j}(\omega_1) = \sum_{k =1}^{N} \mathbf{A}_{jk}(\omega_1) \mathbf{p}_{k}(\omega_1)  \, ,
\end{equation}
once $\alpha_j(\omega_1)$ is known. In our case, we are dealing with small and identical nanoparticles and therefore we can approximate the lineshape of each polarizability $\alpha_{j}$ as a Lorentzian of the form~\cite{zou2004silver}
\begin{equation} \label{Eq:alpha}
\alpha_{j}(\omega_1) = \frac{A_0}{(\omega_{\mathrm{res}}-\omega_1)+\mathrm{i} \gamma} \, , 
\end{equation}
where $A_0$ is a constant, $\omega_{\mathrm{res}}=2\pi c/\lambda_{\mathrm{res}}$ is the center frequency of the LSPR, $c$ is the speed of light and $\gamma$ is the half-width of the LSPR. After solving Eq.~(\ref{Eq:DDA_1}) for $\mathbf{p}_{k}(\omega_1)$, the local field can be calculated using the relationship
\begin{equation} \label{Eq:E_loc_1}
\mathbf{E}(\omega_1) =\epsilon_0^{-1} \alpha^{-1} (\omega_1) \mathbf{p}(\omega_1) \, . 
\end{equation}
In the following sections, this local field acts as one of the fundamental field components driving the nonlinear processes under study. Other local field components $\mathbf{E}(\omega_i)$, oscillating at frequencies $\omega_i$, driving the nonlinear optical processes of interest, can be solved for in a similar manner.

\subsection{Second-order nonlinear processes}
\begin{figure}[htbp]
	\centering
	\includegraphics[width=0.95\linewidth]{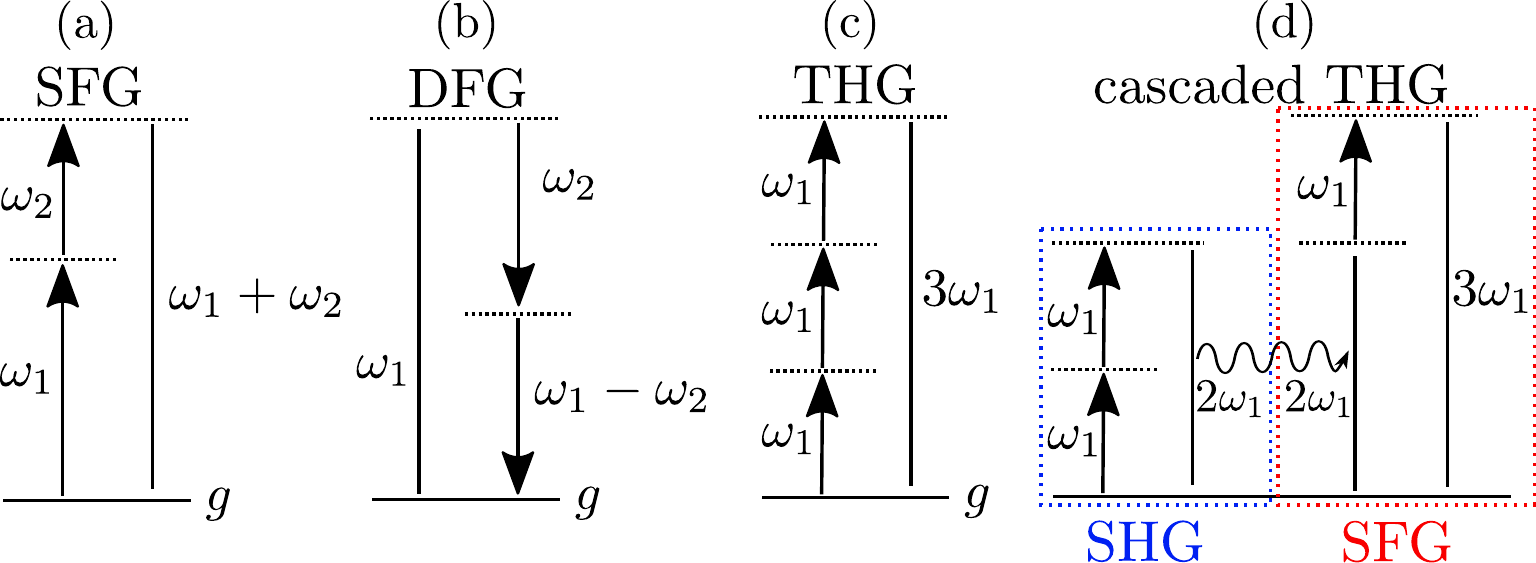} 
	\caption{Schematic energy level diagrams of the studied nonlinear processes. (a) In the process of SFG, an output field oscillating at the frequency $\omega'=\omega_1 + \omega_2$ is created. (b) An output field oscillating at the frequency $\omega'=\omega_1 - \omega_2$ is created in the process of DFG. (c) In the process of direct (non-cascaded) THG, the three incident field components at the frequency $\omega_1$ are converted into the output field oscillating at the tripled frequency $3\omega_1$. (d) Cascaded THG can occur by sequential SHG and SFG processes.}
	\label{Fig:NLO_schematic}
\end{figure}

First, we consider how the second-order nonlinear optical processes are enhanced when the nanoparticles are arranged in periodic arrays. We start by assuming that the undepleted-pump approximation holds, so that the fundamental field is unaffected by the nonlinear processes. We then take the frequency components $\mathbf{E}(\omega_1)$ and  $\mathbf{E}(\omega_2)$ of the local field, solved for as explained above, and use these components to drive the three second-order nonlinear optical processes (SHG, SFG and DFG) occurring at the $j^{\mathrm{th}}$ dipole: 
\begin{subequations}
	\label{Eq:Einc_NLO2} 
	\begin{align}  
	\mathbf{p}_{\mathrm{exc},j}(2\omega_i) = & \epsilon_0 \overset{\text{\small$\leftrightarrow$}}{\beta_j}(2\omega_i ; \omega_i , \omega_i )  \nonumber \\
   & : \mathbf{E}_{j}(\omega_i) \mathbf{E}_{j}(\omega_i) \, , \quad \qquad (\mathrm{SHG})  \label{Eq:Einc_SHG} \\
	\mathbf{p}_{\mathrm{exc},j}(\omega_1+\omega_2) = & \epsilon_0 \overset{\text{\small$\leftrightarrow$}}{\beta_j}(\omega_1+\omega_2 ; \omega_1 , \omega_2 ) \nonumber \\
   & : \mathbf{E}_{j}(\omega_1) \mathbf{E}_{j}(\omega_2) \, , \quad \qquad (\mathrm{SFG})  \label{Eq:Einc_SFG} \\
  	\mathbf{p}_{\mathrm{exc},j}(\omega_1-\omega_2) = & \epsilon_0 \overset{\text{\small$\leftrightarrow$}}{\beta_j}(\omega_1-\omega_2 ; \omega_1 , -\omega_2 ) \nonumber \\
   & : \mathbf{E}_{j}(\omega_1) \mathbf{E}^*_{j}(\omega_2) \, . \quad \qquad (\mathrm{DFG}) \label{Eq:Einc_DFG}
	\end{align}
\end{subequations}
Here the frequencies $2\omega_i$ ($i=$1 or 2), $\omega_1 + \omega_2$ and $\omega_1 - \omega_2$ correspond to SHG, SFG and DFG processes, respectively. The processes of SFG and DFG are schematically shown in Figs.~\ref{Fig:NLO_schematic}(a) and \ref{Fig:NLO_schematic}(b), respectively. The terms $\overset{\text{\small$\leftrightarrow$}}{\beta_j}$ in Eqs.~(\ref{Eq:Einc_NLO2}) are the associated first-order hyperpolarizabilities, and $\mathbf{E}^*_{j}(\omega_2)$ is the complex conjugate of the field $\mathbf{E}_{j}(\omega_2)$. 
One can see by looking at Eqs.~(\ref{Eq:Einc_NLO2}) that the oscillation spectrum of the  $j^{\mathrm{th}}$ dipole moment contains the new frequency components obtained through the nonlinear interactions.

Eqs.~(\ref{Eq:Einc_SHG}) through (\ref{Eq:Einc_DFG}) do not yet take into account the effect the other dipoles and their scattered fields have on the $j^{\mathrm{th}}$ dipole moment, and the calculated dipole moment is thus not yet self-consistent. In order to properly take this effect into account, and thus to find the self-consistent dipole moment components oscillating at these new frequencies, a new system of linear equations needs to be formulated and solved.  
The approach is analogous to the case of the linear response (see Section 2.1). We proceed by defining an excitation field oscillating at the new frequency $\omega'$ as 
\begin{equation}  \label{Eq:E_nlo_exc}
\mathbf{E}_{\mathrm{exc},j}(\omega') = \epsilon_0^{-1} \alpha_j^{-1}(\omega') \mathbf{p}_{\mathrm{exc},j}(\omega')  \, .
\end{equation}
Then we use Eqs.~(\ref{Eq:Einc_NLO2}) and (\ref{Eq:E_nlo_exc}) to calculate the self-consistent dipole moment components $\mathbf{p}(\omega')$ by solving the linear system with $3N$ unknown dipole moment components given by
\begin{equation}  \label{Eq:E_NLO_DDA}
\mathbf{E}_{\mathrm{exc},j}(\omega') = \sum_{k =1}^{N} \mathbf{A}_{jk}(\omega') \mathbf{p}_{k}(\omega')  \, ,
\end{equation}
where the driving term $\mathbf{E}_{\mathrm{exc}}(\omega')$ has the expected quadratic dependence on the incident fundamental field components. We note that, once $\mathbf{p}(\omega')$ has been solved for, it is straightforward to calculate the output field distributions in the far-field locations of interest using, for example, the Green's function formalism~\cite{NovotnyBook}. 

\subsection{Direct and cascaded third-order nonlinear processes}
Next, we calculate how third-order nonlinear optical processes, such as THG or four-wave mixing (FWM), are enhanced in periodic nanoparticle arrays. In addition, we demonstrate how cascading of two second-order nonlinear optical processes can effectively constitute a third-order nonlinear response of an array of nanoparticles. In particular, we show how cascading could be utilized to strongly modify the overall THG efficiency of a given example array. We show this by designing an array exhibiting a strong SLR near the frequency $2\omega_1$ in order to enhance the sequential SHG and SFG processes [see Fig.~\ref{Fig:NLO_schematic}(d)] while there is no SLR near the frequencies $\omega_1$ or $3\omega_1$ present to enhance the direct THG process.

Again, we start by assuming that the undepleted-pump approximation holds. Then we take the local field components $\mathbf{E}_{j}(\omega_i)$, where $i \in \{1,\,2,\,3\}$, obtained through use of Eq.~(\ref{Eq:E_loc_1}), and use these components to obtain the following third-order nonlinear optical processes at the $j^{\mathrm{th}}$ dipole with the dipole moment given by  
\begin{subequations}
\label{Eq:Einc_NLO3} 
\begin{align}  
	\mathbf{p}_{\mathrm{exc},j}(3\omega_1) = & \epsilon_0 \overset{\text{\small$\leftrightarrow$}}{\gamma_j}(3\omega_1 ;\, \omega_1,\,\omega_1,\,\omega_1)  \nonumber \\ 
    &  \vdots \, \mathbf{E}_{j}(\omega_1) \mathbf{E}_{j}(\omega_1) \mathbf{E}_{j}(\omega_1) \, , \quad   ( \mathrm{THG})  \label{Eq:Einc_THG} \\
	\mathbf{p}_{\mathrm{exc},j}(\omega') = & \epsilon_0 \overset{\text{\small$\leftrightarrow$}}{\gamma_j}(\omega'; \,\omega_1,\,\omega_2,\,\omega_3) \nonumber \\ 
    & \vdots \, \mathbf{E}_{j}(\omega_1) \mathbf{E}_{j}(\omega_2) \mathbf{E}_{j}(\omega_3) \, , \quad  ( \mathrm{FWM}) \label{Eq:Einc_FWM} 
	\end{align}
\end{subequations}
where $\overset{\text{\small$\leftrightarrow$}}{\gamma_j}(3\omega_1)$ and $\overset{\text{\small$\leftrightarrow$}}{\gamma_j}(\omega')$ 
are the second-order hyperpolarizabilities giving rise to THG and FWM,  respectively. As previously, the associated excitation field is then calculated using Eqs.~(\ref{Eq:E_nlo_exc}) and (\ref{Eq:Einc_NLO3}). Finally, the self-consistent dipole moment components oscillating at the frequency $\omega'$ due to the occurring third-order nonlinear processes $\mathbf{p}(\omega')$ are calculated by solving a system of linear equations similar to the one of Eq. ~(\ref{Eq:E_NLO_DDA}).
 
As the next step, we proceed to study how cascaded second-order processes contribute to the overall dipole moment component oscillating at $\omega'$. For simplicity, we limit our consideration to the process of THG ($\omega'=3\omega_1$) [see Fig.~\ref{Fig:NLO_schematic}(d)]. We start by calculating the self-consistent dipole moment components oscillating at the frequencies $\omega_1$ and $2\omega_1$ by solving Eqs.~(\ref{Eq:DDA_1}) and (\ref{Eq:E_NLO_DDA}). After this, we calculate the fundamental and the second-harmonic local field frequency components $\mathbf{E}_{j}(\omega_1)$ and $\mathbf{E}_{j}(2\omega_1)$, and use these  components to drive the sequential process of SFG as 
\begin{equation}  \label{Eq:Einc_C_THG}
\mathbf{p}^{\mathrm{C}}_{\mathrm{exc},j}(3\omega_1) = \epsilon_0 \overset{\text{\small$\leftrightarrow$}}{\beta_j} (3\omega_1 ;\, 2\omega_1 , \,\omega_1 ) : \mathbf{E}_{j}(2\omega_1) \mathbf{E}_{j}(\omega_1) \, ,
\end{equation}
where the superscript $\mathrm{C}$ denotes the cascaded contribution to the overall dipole moment component $\mathbf{p}_{\mathrm{tot},j}(3\omega_1)$.
We can now insert this correction to the excitation field oscillating at  $3\omega_1$ due to a direct THG, resulting in a modified system of linear equations, given by
\begin{equation}  \label{Eq:E_THG_total}
\mathbf{E}_{\mathrm{exc},j} + \mathbf{E}^{\mathrm{C}}_{\mathrm{exc},j} = \sum_{k =1}^{N} \mathbf{A}_{jk} \mathbf{p}_{\mathrm{tot},k}  \, ,
\end{equation}
which is then solved to find the total self-consistent dipole moment component $\mathbf{p}_{\mathrm{tot}}(3\omega_1)$. 

\section{Results and Analysis}
We have performed numerical studies with arrays of 201 $ \times $ 201 identical gold nanoparticles arranged in rectangular lattices with lattice periods $p_x$ and $p_y$. The nanoparticles were assumed to be surrounded by a homogeneous medium with a refractive index of $n=1.51$. The linear optical responses of individual nanoparticles were assumed to be well approximated by a Lorentzian lineshape [see Eq.~(\ref{Eq:alpha})]. However, the nonlinear responses of individual nanoparticles were assumed, for simplicity, to be instantaneous and, therefore, dispersionless. This assumption simplifies the interpretation of our results, since the dispersive behavior of the simulated nonlinear optical responses is then solely due to the modified local fields  arising from  the periodic arrangement of the nanoparticles. 

\begin{figure}[ht]
	\centering
	\includegraphics[width=0.55\linewidth]{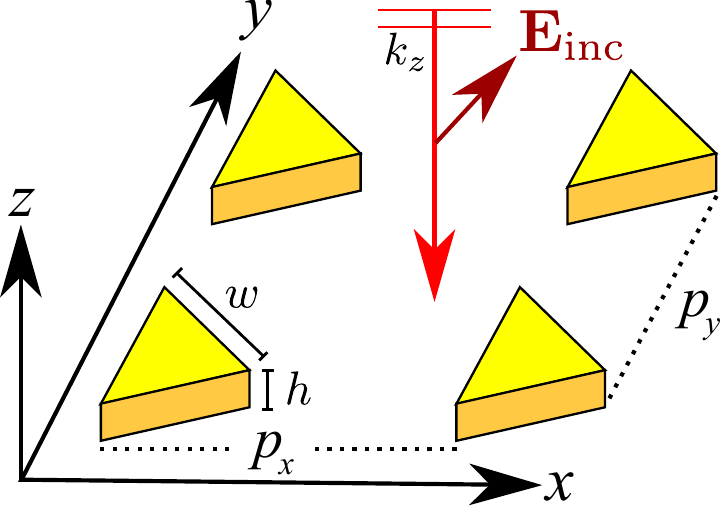} 
	\caption{Incident plane wave propagating along the $z$-direction illuminates a nanoparticle array. The particles are assumed to be equilateral triangular nanoprisms belonging to the symmetry group $D_{3h}$. The side width of the nanoprisms is $w$, and their height is $h$. The particles are arranged into rectangular arrays in the $xy$-plane with array periods $p_x$ and $p_y$. }
	\label{Fig:Sample_schematic}
\end{figure}

For simplicity, we assumed that the arrays were illuminated with plane waves propagating at normal incidence with respect to the surface of the arrays (along the $z$-direction, see Fig.~\ref{Fig:Sample_schematic}). The nanoparticles were assumed to belong to the symmetry group $D_{3h}$, resembling equilateral triangular nanoprisms (see Fig.~\ref{Fig:Sample_schematic}). 
Such symmetry exhibits only one independent non-zero first-order hyperpolarizability tensor component, yielding the hyperpolarizability tensor $\beta_{yyy}=-\beta_{yxx}=-\beta_{xxy}=-\beta_{xyx} = 4.6 \times 10^{-27}\: \rm{m}^4/V $, where the directions $x$ and $y$ lie in the plane of the array. The hyperpolarizability estimate was based on the literature value for an individual gold nanoparticle of similar size~\cite{Butet2010}. 
Similarly, the non-zero second-order hyperpolarizability tensor components are  $\gamma_{ijkl}=\gamma_{xxyy} \delta_{ij} \delta_{kl} + \gamma_{xyxy} \delta_ik \delta_{jl} + \gamma_{xyyx} \delta_{il} \delta_{jl}$, where $[i,\,j,\,k,\,l] \in \{x,\,y \}$. The components related to the $z$-direction were neglected, since our excitation field was purely polarized in the $xy$-plane.  
Due to the symmetry of the nanoparticles, we set $\gamma_{xxyy} = \gamma_{xyxy} = \gamma_{xyyx} = 2 \times 10^{-36} \:\rm{m}^5/V^2$, where the strength estimate was based on the literature values for bulk gold~\cite{FuxianPRB2009, Bloembergen1971}. 
These hyperpolarizability values are simple estimates, and more accurate values can be deduced through appropriate numerical simulations~\cite{Bachelier2008, Zeng2009, Benedetti2010engineering, makitalo2011boundary, Butet2013}, or by measurements~\cite{Butet2010}.
We proceed further with describing the results of our numerical studies of the three nonlinear optical processes, as outlined in the previous section.

\subsection{Difference-frequency generation}
First, we simulated DFG efficiency from an array with periods $p_x=p_y=526$~nm that gives rise to a  first diffraction order (DO) near the wavelength of 794~nm. The nanoparticles in the array were assumed to have the LSPR center wavelength at $\lambda_{\mathrm{res}}=700$~nm. The LSPR center wavelength and the array periods were chosen to strongly enhance the calculated local field components oscillating near the wavelength of 800~nm due to the occurring SLR. The 6-nm redshift of the SLR with respect to the DO has appeared due to the hybridization of the LSPR and DO modes~\cite{Suchowski2017nonlinear}.

The LSPR was assumed to have a Lorentzian lineshape, described by Eq.~(\ref{Eq:alpha}), with $A_0=0.09$~cm$^3 \cdot $s$^{-1}$ and $\gamma=8.3 \times 10^{13}$~s$^{-1}$. These values were found by fitting a Lorentzian lineshape to a simulated extinction spectrum for a single gold nanoprism with the side width $w=60$~nm and the height $h=25$~nm. The extinction simulations were performed using finite-difference time domain (FDTD) method (Lumerical FDTD). The incident plane wave was assumed to be linearly polarized along $x$-direction, while the input wavelength range spanned from 790 to 810~nm. 
  
\begin{figure}[ht]
	\centering
	\includegraphics[width=0.99\linewidth]{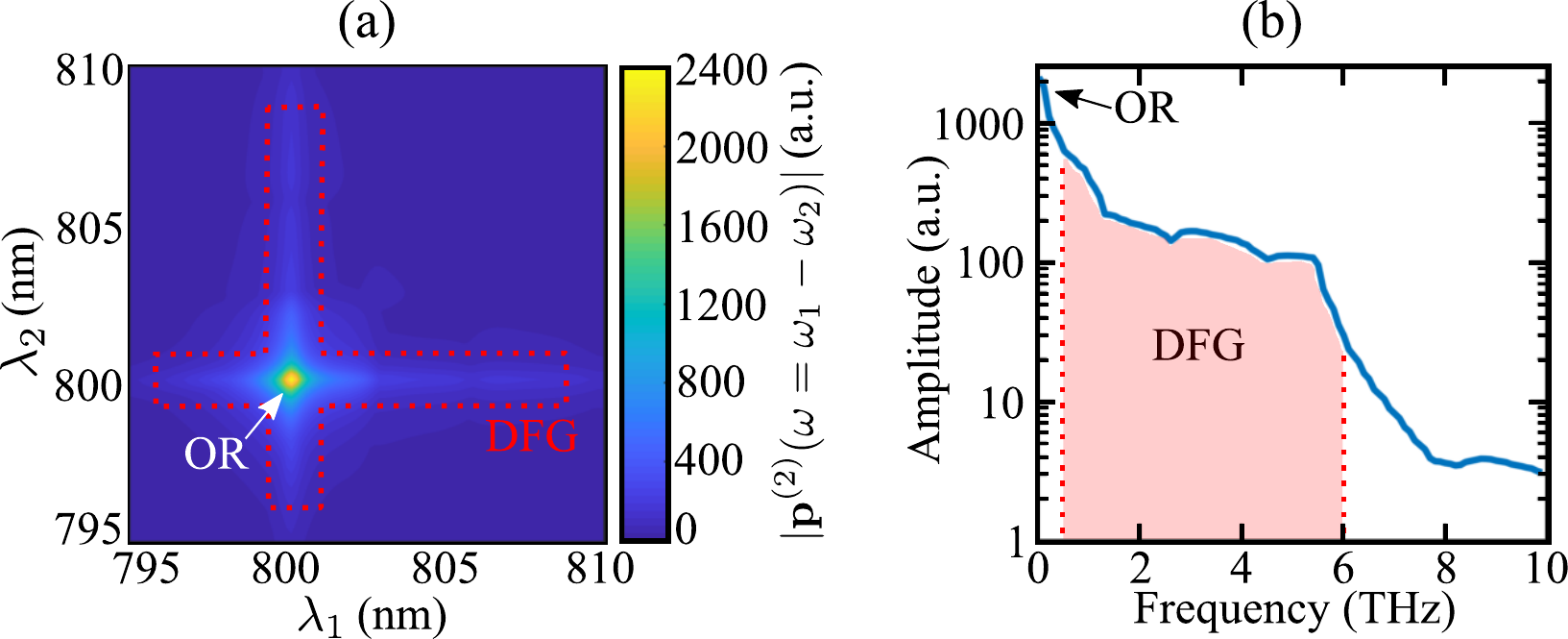}   
	\caption{Nanoparticle array with  periods $p_x=p_y=526$~nm exhibits enhanced DFG/OR response near 800~nm, resulting in THz frequency generation. The results are presented for a single particle at the center of the array. (a) $x$-polarized incident fields with the wavelengths near 800~nm result in an over 2400-fold enhancement in the dipole moment corresponding to the OR (shown with white arrow). The DFG response of the array for the incident fields with the wavelengths near 800~nm is also noticeably enhanced  and exhibits generation of THz radiation (see the area confined by the red dotted lines).  (b) The calculated spectral amplitude of the THz signal field 
exhibits a broadband response between 0.1 and 6 THz with an enhancement factor on the order of  30--1000 in comparison with an individual nanoparticle's response.
}
	\label{Fig:DFG_2D_spectra}
\end{figure}

Using the nonlinear DDA approach, introduced in the previous section, we calculated the DFG dipole moment distribution for the array of nanoparticles. In general, the values of the calculated dipole moments depend on the position of a particle in the array: the largest values occur for the particles at the center of the array. For clarity, the DFG dipole moment components only for the particle at the center of the array are plotted in Fig.~\ref{Fig:DFG_2D_spectra} as functions of the wavelengths of the incident field components. To facilitate the quantitative analysis of the role of the periodic array on the overall DFG efficiency, the results are scaled such that a single and isolated nanoparticle gives a DFG efficiency of unity for any incident wavelength. Therefore, a calculated dipole moment amplitude of 2400 corresponds to a similar enhancement in the amplitude of the generated nonlinear signal field from the particle. 

A more than 2400-fold enhancement of the DFG process has been demonstrated for the incident fields with the wavelengths coinciding with the SLR near 800~nm in the vicinity of the optical rectification (OR) frequency range [see Fig.~\ref{Fig:DFG_2D_spectra}(a)]. The calculated spectral amplitudes of the THz signal field span from 0.1 to 6 THz [see Fig.~\ref{Fig:DFG_2D_spectra}(b)], with an enhancement factor on the order of 30--1000 compared to those of individual nanoparticles. This signal falls inside the well-known terahertz gap, which is defined as the band of frequencies ranging from 0.1 to 10 THz, inside which the generation and detection of radiation is still challenging. Since the earlier works considering plasmonic nanoparticle arrays for THz generation have not utilized SLRs~\cite{Polyushkin2011THz, Luo2014broadband}, we believe that the predicted enhancement factors are very promising. 

\subsection{Sum-frequency generation}
We next simulated the efficiency of SFG from an array of nanoparticles with periods $p_x=759$~nm and $p_y=792$~nm, giving rise to  the first two DOs near the wavelengths of 1146~nm and 1196~nm, respectively. This time, we assumed the particles to exhibit an LSPR with the center wavelength $\lambda_{\mathrm{res}}=1000$~nm, and to be well approximated by a Lorentzian lineshape  [see Eq. (\ref{Eq:alpha})] with $A_0=0.43$~cm$^3 \cdot$s$^{-1}$ and $\gamma=13.8 \times 10^{13}$~s$^{-1}$. These values were found by fitting a Lorentzian lineshape to a simulated extinction spectrum of a single gold nanoprism with the dimensions $w=210$~nm and $h=25$~nm (FDTD Lumerical). 
This array configuration results in two non-degenerate SLRs formed near 1160~nm and 1205~nm, associated with the array periods $p_x$ and $p_y$, respectively.  

\begin{figure}[ht]
	\centering
	\includegraphics[width=0.99\linewidth]{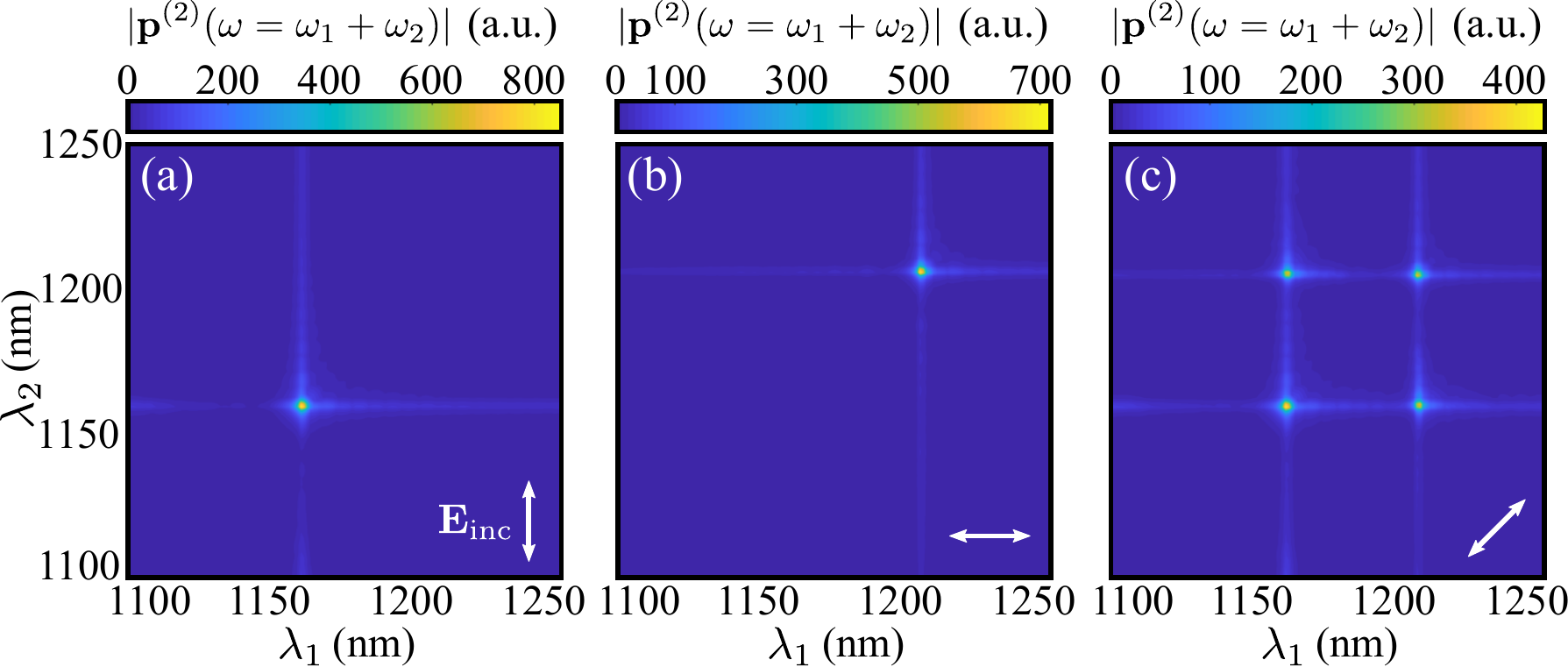} 
	\caption{Nanoparticle array with  periods $p_x=759$~nm and $p_y=792$~nm, giving rise to SLRs near the wavelengths of 1160~nm and 1205~nm, respectively, exhibits enhanced SFG/SHG response when the wavelengths of the incident fields coincide with the SLRs. We plot the SFG dipole moment  amplitude for the particle at the center of the array for the incident fields, linearly polarized along (a) $y$-, (b) $x$- and (c) $(x+y)$-direction. The dipole moment is scaled with respect to that of a single isolated particle.}
	\label{Fig:SFG_2D_spectra}
\end{figure}

Linearly polarized incident fields were used for excitation with the wavelengths ranging from 1100 to 1250~nm. We considered three different linear polarizations of the incident field, oriented along $y$-, $x$- and $(x+y)$-direction. The latter polarization orientation corresponds to the linear polarization rotated at 45~degrees with respect to the $x$-orientation towards the $y$-direction (see Fig.~\ref{Fig:Sample_schematic}). This analysis allowed us to study the dependence of the excitation of SLRs on the polarization of the incident light. We calculated the SFG dipole moment of the particle at the center of the array and plot it as the function of the two incident fields' wavelengths. The results are displayed in Fig.~\ref{Fig:SFG_2D_spectra} for the three incident field polarizations, as described above. When the incident field is polarized along $y$-direction ($x$-direction), it is effectively coupled with the SLR at 1160~nm (1205~nm), as shown in Fig.~\ref{Fig:SFG_2D_spectra}(a)  [Fig.~\ref{Fig:SFG_2D_spectra}(b)]. When the input polarization is oriented along  $(x+y)$-direction, both SLRs are simultaneously excited [see Fig.~\ref{Fig:SFG_2D_spectra}(c)]. These findings demonstrate that one can engineer the efficiency of the nonlinear optical responses, and can control it by changing the incident polarization. 

\subsection{Direct and cascaded third-harmonic generation}
Finally, we studied how the third-order nonlinear optical processes can be enhanced through engineering the nanoparticle arrays. In particular, we determined whether (and how) the two sequential second-order nonlinear optical processes of SHG and SFG can contribute to the overall THG efficiency [see the schematics of the associated processes in Figs.~\ref{Fig:NLO_schematic}(c)~and~\ref{Fig:NLO_schematic}(d)]. We assumed the wavelength of the incident field to vary between 1500 and 1600~nm. In order to enhance the effect due to cascading of the two second-order nonlinear optical processes [see Fig.~\ref{Fig:NLO_schematic}(d)], we designed an array where the local field is strongly enhanced by an SLR near 770~nm. The LSPR center wavelength was chosen to be at 700~nm, which was achieved by using the particles identical to those used for the DFG studies. Then, as a consequence of arranging the particles into an array with periods $p_x=p_y=505$~nm, a DO near the wavelength of 763~nm appeared, giving rise to the SLR near 770~nm. 

\begin{figure}[ht]
	\centering
	\includegraphics[width=0.95\linewidth]{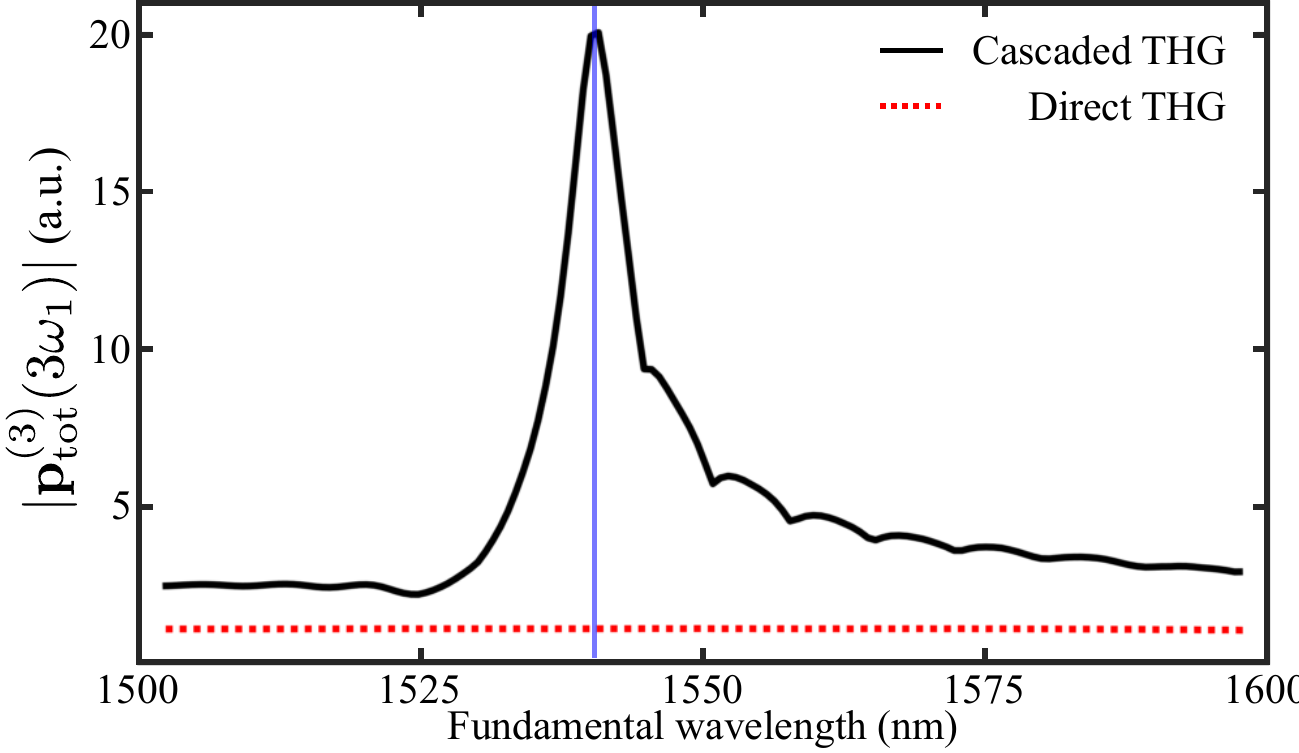} 
	\caption{Calculated THG dipole moment amplitude for the nanoparticle at the center of the array with periods $p_x=p_y=505$~nm, giving rise to a SLR near the wavelength of 770~nm. When the cascaded contribution is neglected (see the red dotted line), the THG response of the array does not markedly depend on the incident fundamental wavelength because there is no strong resonances either near the fundamental wavelength (1500--1600~nm) or near the THG signal (500--533~nm). When the cascaded nonlinear effects are taken into account (see the black solid line), a clear peak near the fundamental wavelength of 1540~nm (highlighted with the vertical blue line) appears, demonstrating a 20-fold enhancement of the overall dipole moment component corresponding to THG.}
	\label{Fig:THG_spectra}
\end{figure}

For an incident field oscillating near 1540~nm, the SHG process is enhanced by the SLR near 770~nm, while no resonance exists either near the fundamental (1540~nm) or near the THG wavelength (513.3~nm). Therefore, the overall enhancement of the THG efficiency for the fundamental wavelength 1540~nm can be fully attributed to the cascaded process. The results of the THG simulations with the cascaded effects both taken into account and neglected are shown in Fig.~\ref{Fig:THG_spectra}. In the simulations, the incident field was assumed to be linearly polarized along $x$-direction. The black solid like shows the normalized dipole moment for the THG process with both the contributions (direct and cascaded) tken into account, while the dashed red line shows the dipole moment for the direct process only. As expected, the overall efficiency of THG is resonantly enhanced due to a strong cascaded contribution arising from the sequential SHG and SFG processes.
  
\section{Discussions and Conclusion}
We next discuss the potential of the demonstrated nonlinear DDA approach to be used for better understanding the nonlinear optical processes in nanoparticle arrays. The proposed method is based on recently introduced nonlinear DDA approach to simulate the optical response of individual particles~\cite{balla2010second}, where the resulting systems of linear equations are solved iteratively using generalized minimal residual method and custom-written Matlab codes. The block-toeplitz nature of the interaction matrix $A_{jk}$ and fast Fourier transforms were utilized to reduce the memory constraints and to speed-up the computations~\cite{goodman1991application, draine1994discrete}. The calculations were further accelerated by performing the most demanding computations using a graphics processing unit (GeForce GTX TITAN X) resulting in an additional $\sim$10-fold speed-up. We note that similar results could be achieved using other computational approaches, for example, based on FDTD~\cite{Joseph1997} or finite-element methods~\cite{makitalo2011boundary, Benedetti2010engineering, Butet2013}. We chose the nonlinear DDA approach, since it also provided an intuitive understanding of the underlying physics and was straightforward to implement. In addition, the anisotropic and tensorial optical responses would be straightforward to implement using the DDA approach, which are known to often play a major role in the nonlinear responses of individual nanoparticles~\cite{czaplicki2013enhancement}. The  DDA approach could also be a useful tool for studying how fabrication imperfections or deviations of the array geometry from the designed one could affect the predicted optical response of the array. Utilizing more commonly used simulation tools, such as the FDTD method, for such studies would be significantly more demanding computationally because periodic boundary conditions cannot be applied to simplify such simulations.

In conclusion, we have studied the nonlinear responses of plasmonic nanoparticle arrays. In particular, we studied how the collective responses of nanoparticle arrays known as surface lattice resonances could be utilized to enhance and engineer nonlinear processes such as second-harmonic, sum-frequency, difference-frequency and third-harmonic generation. We reported on significant enhancements (20- to 2400-fold) of nonlinear optical responses, which could be useful, for example, to achieve efficient terahertz generation from nanoparticle arrays. We have also studied how surface lattice resonances could be utilized to enhance higher-order nonlinear optical processes through cascaded lower-order processes, with the implications of generation of ultraviolet radiation using nanoparticle arrays. As an example, we have shown, using a simple array design, that the process of third-harmonic generation could be enhanced almost 20-fold due to cascaded second-order nonlinearities. We believe that our results could be useful for various applications where nonlinear phenomena are utilized, including novel sources of coherent light.
 
\section*{Funding}
Finnish Cultural Foundation (00150020); the Academy of Finland (308596); the Canada Excellence Research  Chairs Program; the Canada Research Chairs Program; the  Natural Sciences and Engineering Research Council of Canada (NSERC) Discovery funding program.


\end{document}